\documentclass[12pt,preprint]{aastex}

\shorttitle{Wind-clumping in SMC WR stars}
\shortauthors{Marchenko et al.}

\begin{document}

\title{Spectroscopy of SMC Wolf-Rayet Stars Suggests that Wind-Clumping does not Depend on Ambient
Metallicity}

\author{S.V. Marchenko\altaffilmark{1},  C. Foellmi\altaffilmark{2,3},  A.F.J. Moffat\altaffilmark{4},
F. Martins\altaffilmark{5}, J.-C. Bouret\altaffilmark{6}, E. Depagne\altaffilmark{3,7}}

\altaffiltext{1}{Department of Physics and Astronomy, Western Kentucky
University, 1906 College Heights Blvd., 11077, Bowling Green, KY
42101-1077; sergey.marchenko@wku.edu}
\altaffiltext{2}{Laboratoire d'Astrophysique, Observatoire de Grenoble, BP 53,
38041 Grenoble Cedex 9, France; cfoellmi@eso.org}
\altaffiltext{3}{European Southern Observatory, 3107 Alonso de Cordova, Casilla 19001,
Vitacura, Santiago, Chile}
\altaffiltext{4}{D\'epartement de Physique and
Observatoire du Mont M\'egantic, Universit\'e de Montr\'eal, CP 6128,
Succursale Centre-Ville, Montr\'eal, QC H3C 3J7, Canada; moffat@astro.umontreal.ca}
\altaffiltext{5}{Max-Planck-Institut f\"{u}r extraterrestrische Physik, Postfach 1312,
D-85741 Garching,  Germany; martins@mpe.mpg.de}
\altaffiltext{6}{Laboratoire d'Astrophysique de Marseille, Traverse du Siphon - BP 8,
13376 Marseille Cedex 12, France; jean-claude.bouret@oamp.fr}
\altaffiltext{7}{Departamento de Astronomía y Astrof\'{i}sica, Pontificia Universidad
Cat\`{o}lica de Chile, Campus San Joaqu\'{i}n, Vicu\~{n}a Mackenna 4860 Casilla 306,
Santiago 22, Chile; edepagne@eso.org}

\begin{abstract}
The mass-loss rates of hot, massive, luminous stars are considered  a decisive
parameter in shaping the evolutionary tracks of such stars and influencing the interstellar
medium on galactic scales. The small-scale structures (clumps) omnipresent in
such winds  may reduce empirical estimates of mass-loss rates by an evolutionarily significant factor
of $\geq 3$. So far, there has been no {\it {direct}}  observational evidence that
wind-clumping may persist  at the same level  in environments with a low
ambient metallicity, where the wind-driving opacity is reduced.  Here we report
the results of time-resolved spectroscopy of three presumably single Population I Wolf-Rayet
stars in the Small Magellanic Cloud, where the ambient metallicity is $\sim 1/5 Z_\sun$.
We detect numerous
small-scale emission peaks moving outwards in the accelerating parts of the stellar winds.
The general properties of the moving features, such as their velocity dispersions,
emissivities and average accelerations, closely match the corresponding characteristics
of small-scale inhomogeneities in the winds of Galactic Wolf-Rayet stars.

\end{abstract}


\section{Introduction}

Wolf-Rayet (W-R) stars, as a final evolutionary phase of hot, massive stars
($M_{init} \geq 25 M_{\sun} $ for a $Z\sim Z_{\sun} $ ambient metallicity), are especially sensitive
to all phenomena which may influence their mass-loss rates. The fast, dense
winds of W-R stars are driven by radiation pressure exerted on multiple lines of mainly heavy
elements. Hence, W-R mass-loss rates should be sensitive to the
ambient metallicity content, as well as to the locally enhanced chemical composition of the wind.
The former sensitivity, though suggested from general principles
(Lamers \& Cassinelli 1999), has escaped detection until recently (see Crowther 2006,
Gr{\"a}fener \& Hamann  2006, and references therein).

Line-driven winds are inherently unstable (Lucy \& Solomon 1970), being
constantly fragmented by numerous embedded shocks (Dessart
\& Owocki 2005). Such fragmentation may
change the local ionization balance and create a local non-monotonic velocity field,
thus providing a strong feedback to the driving force.
In the past, wind diagnostics
relied on models of smooth, homogeneous winds, until numerous spectroscopic
observations of Galactic W-R and OB stars
(Moffat et al. 1988; Robert 1992; L\'epine \& Moffat 1999; Eversberg et al. 1998;
Bouret et al. 2005; Fullerton et al. 2006 and references therein) demonstrated
the omnipresence of wind-embedded clumps, usually taking on the form of discrete
density enhancements outmoving with an accelerating wind. As an immediate
and long-lasting impact, the relatively simplistic treatment of structures in the
otherwise highly elaborate models of hot-star winds
(Hillier \& Miller, 1999; Puls et al. 2005; Hamann et al. 2006 ) resulted in a
consensual downward revision of mass-loss rates by  a factor of  $\geq 3$,
confirmed by numerous independent observations (e.g.,  Moffat \& Robert 1994).

Though wind-clumping was widely anticipated to operate also in low-Z environments
(Hamann \& Koesterke 1998; Crowther
et al. 2002; Bouret et al. 2003), any {\it {direct}} proof was lacking. Here we report the detection of
outmoving spectral features in the line profiles of 3 SMC W-R stars. We compare the general
characteristics of these clumps to those in a Galactic sample of W-R stars, finding them to be
strikingly similar.

\section{Observations}

We targeted three presumably single (Foellmi et al., 2003) early-type WN stars in the SMC:
SMC-WR 1 (WN3ha), WR 2 (WN5ha) and WR 4 (WN6h), thus forming a representative sample of the small SMC W-R population
(12 known Population I W-R stars, most of them early-type WNs: Massey et al. 2003).
We monitored the stars in continuous, $\sim$1h 40 min-long loops for two consecutive
nights in August, 2006 (between HJD 2,453,974.538 and ...75.877), alternating among the 3 targets:
 WR 4 - WR 2 - WR 1... .
We used the  UVES spectrograph at the ESO-VLT-UT2 8-m telescope (Kueyen),
sampling the region $\lambda 3927-6031$\AA\ .
During the routine reduction we experimented  with different binning factors, finding a
reasonable compromise between
desired spectral resolution, $\lesssim 0.5$\AA\ /pix, and signal-to-noise ratio,  $S/N> 100$, by binning
the available spectra to 0.25\AA\  (the region around  HeII 4686\AA\ ) and  0.31\AA\  (HeII 5412\AA\ ) bins.
With 40, 30 and 20 min exposures, this resulted in 5-6
spectra/night/star, with comparable signal-to-noise ratios
$S/N \simeq 120$, $\simeq 140$
and $S\simeq 160$ ($\pm 20$) for WR1, WR2 and WR4, respectively,
estimated from the adjacent continua around measured spectral lines.

\section{Properties of the Clumps}

The high quality of the spectra allowed us to apply a rather straightforward procedure to
detect the line-profile variability. We rectified the profiles of all relatively prominent HeII
lines (the only lines strong enough to provide sufficient precision, namely,
HeII 4686, 4859 and 5412 \AA\ ) by linear interpolation between the line-free regions
located in the immediate vicinity of a given line. Then we created unweighted average profiles
of each line for each night of observation and subtracted these averages from the individual
profiles. We found that all the major emission lines in all 3 stars can be considered as variable at a
level substantially exceeding the predicted instrumental noise level.

For a proper comparison of the new results with previous studies of
inhomogeneities in the winds of Galactic W-R stars, we closely follow the already developed
approaches specifically designed to reveal basic properties of clumps.

(i) First of all, we attempt to detect similarities in the variability patterns of major emission lines
by cross-correlating  the difference (individual - average) spectra
of different line transitions. We use
the average cross-correlation functions (see L\'epine et al. 2000, equations 7, 8): namely,
we produce an unweighted average of the
individual (transition-to-transition) cross-correlation functions for a given star and a given spectral window
(usually, $\pm v_\infty$, centered on the transition's $\lambda _0+\Delta \lambda$, where $\Delta \lambda$
corresponds to the SMC's systemic velocity).  The maxima of the
average cross-correlation functions reach 0.2-0.4, which is comparable to $I_{max}=0.2-0.8$ measured in the
CIII and CIV profiles of the Galactic star
WR 135. Taking into account
that (a) the considered transitions arise from different, though overlapping, line-forming regions,
(b) there is some hydrogen present in the $\lambda 4860$\AA\ blend, and (c) there might be some
transition-dependent optical-depth effects (see below), we did not anticipate a very high
degree of correlation. Nevertheless, for general consistency, plus a slight gain
in S/N, we combine the appropriately shifted (to the rest-frame of
the HeII 4686 transition) difference spectra of the three lines, smooth them with a
gaussian filter (FWHM=7 pix), linearly  time-interpolate the neighboring spectra in order to fill
the unavoidable gaps, and plot the results in Fig. 1. The continuous, ${\bold {\rm V}}$-shaped
structures present in the difference spectra of the SMC W-R stars are remarkably
similar to the variations
seen in {\it all} appropriately studied Galactic WR stars (L\'epine \& Moffat 1999),
with the latter firmly linked to the presence of numerous
small-scale inhomogeneities moving outwards in the accelerating parts of the stellar winds.

(ii) We proceed with measurements of the clump properties closely following the approaches
developed in L\'epine \& Moffat (1999) and  in L\'epine et al. (2000). Namely, we use the
degradation function to estimate the average acceleration of the outmoving emission features
(Fig. 2) and the net intrinsic variability levels, $\sigma_{intr}$,
in different transitions (Fig. 3). We remind the reader that the degradation function
(L\'epine et al. 2000, equation 2) measures the mean standard deviation between
spectra separated by a [variable] interval $\Delta t$. In addition, this function also accounts for a possible
motion of the clumps (predominantly, away from the line's center), by including the  average acceleration
as a free parameter. The levels of net intrinsic variability
are estimated  across emission-line profiles (L\'epine et al. 2000, equations 4,5) by using the
variability in line-free continuum regions adjacent to emission lines and treating it as a
source of  instrumental noise,
then adjusting this noise level by allowing for the difference in count statistics across the profile of
a prominent emission line.

(iii) In order to compare our measurements of the individual
fluxes and full-widths at half-maxima of the detected clumps with  corresponding values
for Galactic WR stars, we follow the general  approach and published measurements from Robert
(1992), thus using only the HeII 5412\AA\ line. Namely, we measure the fluxes, maximum intensities
and FWHMs from the least-square fits of Gaussian profiles to the recognizable features in the
difference-spectra (individual spectrum - night average).
Fig. 2 shows that, within the
statistical uncertainties, which are larger for the SMC stars mainly due to the relatively lower
S/N, the properties of the Galactic clumps are indistinguishable from those in the SMC sample.
Note that for the estimates of the clump properties we use the unsmoothed difference-spectra.

The broad spectral coverage of the data allows us to address the question of optical depth of the
clumps by comparing the intrinsic variability patterns in different transitions. One should note
(Fig. 3) the most instructive contrast between HeII 4686\AA\  (transition 3-4) and  HeII 5412\AA\
(transition 4-7). In order
to make quantitative inter-comparisons of the intrinsic variability levels, we normalise $\sigma_{intr}$
by the line-profile intensity of the corresponding transition.
Both SMC-WR 1 and WR 2 show a reasonably consistent behavior in both transitions (Fig. 3;
compare to Fig. 6 from L\'epine et al. 2000), if
one takes into account the pronounced difference in the extensions of  the line-forming regions: compare
the velocity extensions of  the average profiles of HeII 4686\AA\  to their  HeII 4859 and  5412\AA\ counterparts.
Surprisingly, the variability patterns of WR 4 come as a sharp contrast to the predictably congruent
behavior of the patterns in WR 1 and WR 2.

\section {Discussion and Conclusions}

First of all, we concentrate on the intrinsic line variability as related
to optical-depth effects introduced by the clumps. Relying on the differences in the
total optical depths of the studied transitions and assuming that these are  also applicable to the
population of discrete density enhancements, one expects to detect different
$\sigma_{intr}$ dependencies across the profiles. Most notably, the significant optical
depth of HeII 4686\AA\ may result in a blue-shifted peak of $\sigma_{intr}$  relative to
HeII 4859\AA\ and HeII 5412\AA\ . Quite predictably, SMC-WR 1 and
WR 2 show little, if any, line-opacity effects: note the pronounced (Fig. 3) absorption components
in the profiles of HeII 4859\AA\ and HeII 5412\AA\ , which are generally linked to low
optical depths of the winds. Concurrently, L\'epine \& Moffat (1999) find only mild clump-related
optical-depth effects in the [optically thin] CIII 5696\AA\ line formed in the winds of
Galactic WC stars, along with the possible presence of opacity-related effects in HeII 5412\AA\
seen in the Galactic WN stars.
Hence, for WR 1 and WR 2 one may conclude that either (a) the clumps are optically thick in all
transitions {\it and} the winds have rather low volume filling factors
(e.g., Owocki \& Cohen 2006), or (b) all clumps are optically thin. Unfortunately, our present limited
data set does not allow one to choose between these two alternatives.
The only noticeable difference between WR 1 and WR 2 is the secondary maximum
in $\sigma_{intr}$ of HeII 4859\AA\ for WR 2, which may be caused by the presence of relatively high quantities of hydrogen in the wind.

WR 4 tells a different story. The intensity of the HeII 4686\AA\ transition, as well as the lack of
developed P Cygni absorptions, calls for a denser wind. If the global properties of the wind are
related to the properties of clumps, then one may detect different  wavelength
dependencies of $\sigma_{intr}$. This clearly pertains to WR 4, where the blueward displacement
of the HeII 4686\AA\ peak may point to a high optical density of clumps. However, the
disparity of the HeII 4859 and 5412\AA\ distributions prompts an alternative explanation:
the spectral variability could be, in part, related to the presence of co-rotating interacting
regions (CIRs: Massa et al. 1995; Morel et al. 1999, and references therein), as seen in several
Galactic W-R stars. One may notice
the synchronized {\it redward} migration of weak emission features in the first half of Night 1 (Fig. 1, dashed
lines: $\lambda \sim 4673$ shifting to $\lambda \sim 4676$,  $\sim 4689$ gradually moving to $\sim 4693$,
and $\sim 4698$ moving to $\sim 4702$), as well as the gradual disappearance of the
broad emission feature at $\lambda \sim 4690$ during Night 2, both phenomena reminiscent of a
CIR. In addition, along with general expectations, WR 4 is a periodic photometric variable
(Foellmi et al. 2003: P=6.55 d), and not an obvious binary.  This period is then likely related to rotation.
However, this does not mean that WR4 lacks clumps.

One should also notice the lack of any dependence of clump
FWHMs  on spectral class or terminal
velocity of the W-R winds (Fig. 2, lower section). The recent comprehensive
analysis of turbulence in colliding supersonic flows (Folini \& Walder 2006)
shows that the velocity dispersions in the shock-bound cold dense layers
(the clumps?) depend only on the upwind Mach-numbers, $M_u$.
We consider a  `generic' W-R wind with $v_\infty=2000$ km/s,
estimating the dispersion of turbulent velocities as $\sigma(v_{turb}) \sim (0.1-0.2) v_\infty$
(Prinja et al. 1990; Marchenko \& Moffat 1999).
We assume an `asymptotic' wind temperature $T=2\cdot10^4$K. This provides
an estimate for the sound speed in the clump-formation zone,
$\sim 5-100 R_\star$ (see the models in L\'epine et al. 2000).
Using  $M_{rms} \sim \eta^{-1/2} M_u$ with
density contrast
$\eta \sim 30$ (Folini \& Walder 2006), one may estimate the typical width of a clump as:
$FWHM \sim (1.4-2.7)$\AA\ , or 90-170 km/s at $\lambda = 4686$\AA\ . This rough estimate
falls tantalizingly close
to the  measurements from Fig. 3, which give  $FWHM=100-300$ km/s. One may match these
figures even more closely under
the assumption that the hydrodynamically-modeled density contrasts fall  below the pre-specified
value of $\eta=30$, by adopting $\eta\sim10$ as frequently used in models of W-R spectra
(Hamann \&  Koesterke 1998).

Hence, quite justifiably, we favor the notion in which the clumps have
the $\sim$same internal velocity dispersion defined by compressible
turbulence in a supersonic flow, but different emissivity volumes
(cf. L\'epine \& Moffat 1999).

Concluding our study, we can state that:

\hskip -0.8cm (a) despite the differences in ambient metallicity, both the winds of Galactic and SMC
W-R stars show a clear presence of small-scale structures;\\
(b) the general properties of these structures, namely average acceleration rate, average
 individual flux and average velocity dispersion inside the clump, are similar for
 both populations of W-R stars;\\
(c) point (b) comes as a big surprise, considering that the wind-driving force (and thus
the radiatively-driven instability) sensitively depends
on the presence of heavy elements (Vink \& de Koter 2005);\\
(d) a wind-clumping factor must be included in the models of massive-star winds
in  low-Z environments, just as in high-Z environments. Our understanding of the first population of stars in the
early Universe, mostly all very massive, as well as processes of enrichment of the interstellar
medium, will depend on reliable estimates of mass-loss rates, i.e., the
assumed wind-clumping factors. Indeed, wind-clumping could be rather extreme in a low-Z
environment: one may recall f=0.01 used in the models of MPG 324 (O4 V((f)) star in SMC: Bouret et al. 2003),
and f=0.06 for Br43 (WC4 star in LMC: Hamann \& Koesterke 1998), to be contrasted with the more modest  $f\sim 0.1-0.3$
values for Galactic WR stars. However, note that the recent data on Galactic OB stars
also provide $f \ll 0.1$ (Bouret et al. 2005; Fullerton et al. 2006). This may signify that: either the
filling factors depend on the evolutionary status of massive stars, or the invoked wind-clumping factors
are biased by the model approach and the choice of diagnostic lines. The latter, in turn,
may be related to a [strong] radial dependence of the wind-clumping factor. This obviously calls for a
thorough investigation which is far beyond the scope of this paper.

The extreme clumping may dramatically decrease
estimates of the mass loss rates, thus helping to retain the expected high rotational velocities
 for the progenitors of the
long/soft $\gamma$-ray bursts, as well as affect the appearance of their immediate environments,
thus the observable reaction (echo) on the incoming burst.

\acknowledgments
This paper is based on
observations made with ESO Telescopes at the Paranal Observatory
under programme ID 077.D-0029.
C.F. thanks C. Esparza for the efficient and friendly technical support during observations.
AFJM is grateful for financial assistance to NSERC (Canada) and FQRNT (Qu\'ebec).
FM acknowledges support from the Alexander von Humboldt foundation.

\clearpage

\begin{figure}
\plotone{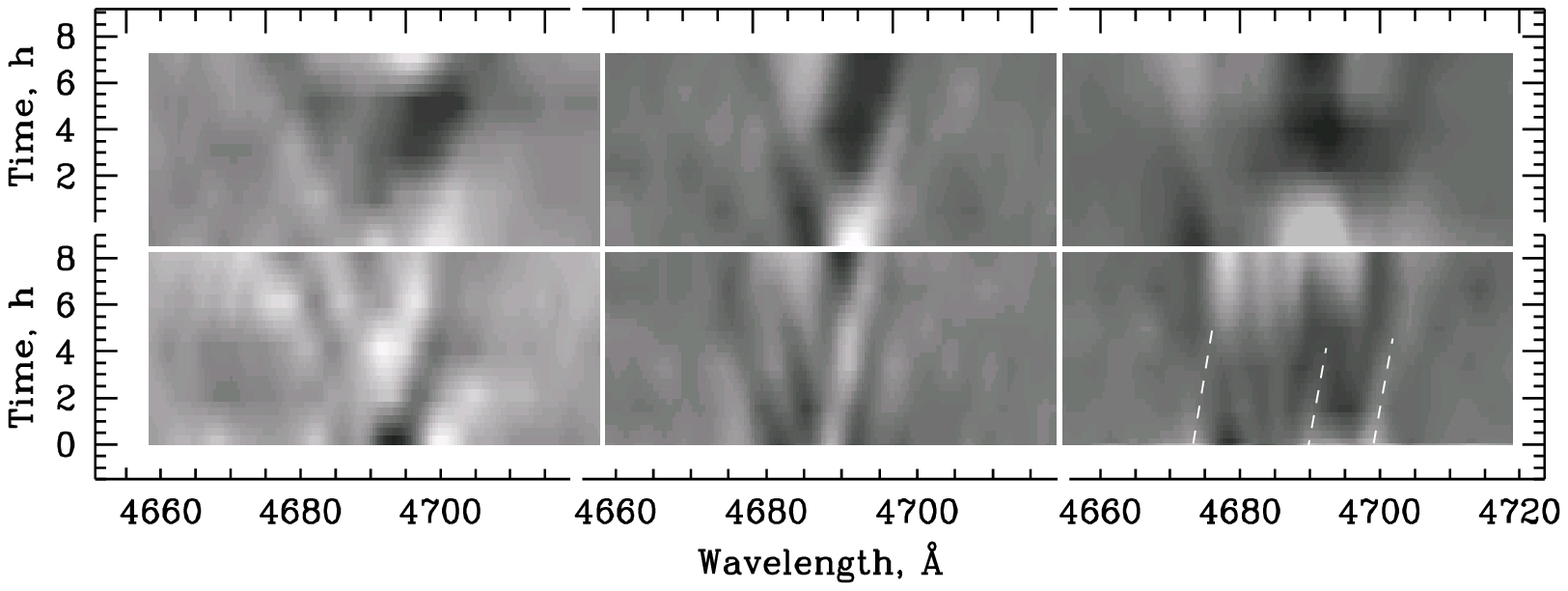}
\caption{Grayscale plots of time-interpolated and smoothed
difference (individual - night-average) spectra of WR1 (left panels),
WR2 (central panels) and WR4 (right panels) for Night 1 (lower panels)
and Night 2 (upper panels).  The intensity ranges are -0.03 (black) to +0.03
(white) in the local continuum units ($\equiv 1$). The dashed lines in the lower-right panel
mark the features presumably related to CIRs (see text). \label{fig1}}
\end{figure}

\clearpage

\begin{figure}
\epsscale{0.35}
\plotone{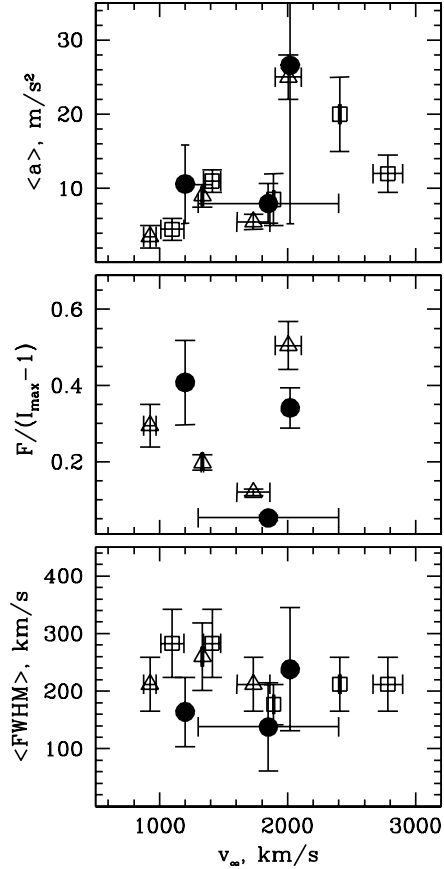}
\caption{The upper panel shows the average
accelerations of outmoving clumps in the winds
for SMC stars (filled circles),  as well as  the measured ranges for Galactic WC
(open squares) and WN (open triangles) stars (see L\'epine \& Moffat 1999).
The data are arranged by the corresponding
terminal velocities of the WR winds.
The middle panel shows the average fluxes
($F$) of the clumps detected in the HeII 5412\AA\ line and  normalized by
the maximum intensity of this emission profile ($I_{max} -1$), in order to be compared
with the measurements of Robert (1992).
The lower panel shows the average
FWHMs of the clumps observed in the HeII 5412\AA\ profile.
Note that WR 134 was omitted from the lower panel due to the very
high $FWHM\sim 800 km/s$ (see explanations in L\'epine \& Moffat 1999).
The $v_\infty$ estimates are rather uncertain for WR 1 and WR 2 (P. Crowther,
priv. comm.); for WR 4 they come from Crowther (2000) and Willis et al.
(2004). \label{fig2}}
\end{figure}

\clearpage

\begin{figure}
\epsscale{1.0}
\plotone{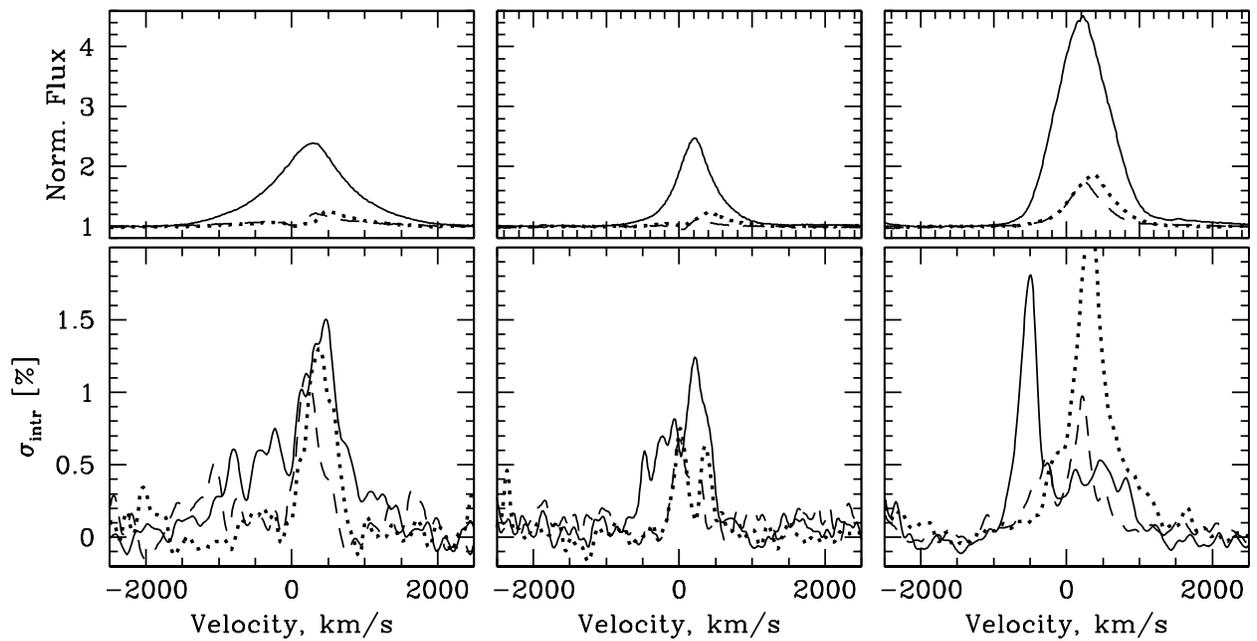}
\caption{The intrinsic variability across the line profiles in WR 1, WR2 and WR 4 (lower panels, left-to-right):
full lines correspond to the HeII 4686\AA\ line, dotted lines show HeII+H 4860\AA\ , dashed lines trace
HeII 5412\AA\ . The upper panels show average rectified profiles of the transitions. \label{fig2}}
\end{figure}

\end{document}